# Knowledge-based system for collaborative process specification

Vatcharaphun Rajsiri*, Jean-Pierre Lorré*, Fréderick Bénaben**, Hervé Pingaud**

* EBM WebSourcing

10 Avenue de l'Europe, 31520 Ramonville Saint-Agne, France

** Centre de Génie Industriel, Université de Toulouse-Mines d'Albi,

Campus Jarlard, Route de Teillet, 81013 Albi Cedex 09

**Abstract**

This paper presents an ontology-based approach for the design of a collaborative business process model (CBP). This CBP is considered as a specification of needs in order to build a collaboration information system (CIS) for a network of organisations. The study is a part of a model driven engineering approach of the CIS in a specific enterprise interoperability framework that will be summarised. An adaptation of the Business Process Modeling Notation (BPMN) is used to represent the CBP model. We develop a knowledge-based system (KbS) which is composed of three main parts: knowledge gathering, knowledge representation and reasoning, and collaborative business process modelling. The first part starts from a high abstraction level where knowledge from business partners is captured. A collaboration ontology is defined in order to provide a structure to store and use the knowledge captured. In parallel, we try to reuse generic existing knowledge about business processes from the MIT Process Handbook repository. This results in a collaboration process ontology that is also described. A set of rules is defined in order to extract knowledge about fragments of the CBP model from the two previous ontologies. These fragments are finally assembled in the third part of the KbS. A prototype of the KbS has been developed in order to implement and support this approach. The prototype is a computer-aided design tool of the CBP. In this paper, we will present the theoretical aspects of each part of this KbS as well as the tools that we developed and used in order to support its functionalities.

## 1. Introduction

Enterprises are nowadays operating in a complex economical environment where markets are more open and globalised. By this way, their competitiveness shall be intensively improved and changes in business models are necessary. As a concrete consequence of these new requirements, modern types of industrial networks have been set up. This fact is widely recognized as a major innovation in business management. It is obviously one of the most important trends in industrial engineering practices for the two last decades and is naturally becoming a consideration subject for numerous research studies. The analysis of business practices in cooperative organizations delivers new corpus of knowledge. For example, a large characterization and classification study of collaborative networked organizations is proposed in [1]. Core competencies and extensive cooperation could be identified as main sources for business efficiency expected by such new forms of organization whatever their nature. Consequently, the capacity of one enterprise to join a target collaborative networked organization, to adapt and react rapidly to market dynamics, as well as the synergy developed by such networked organizations, are subjects of prime interest that address the problem of enterprise architectures with a new point of view.

Enterprise integration is a major issue for information system design (IS) when business performance improvement is expected. Particularly, the quality of communications between actors and systems is deeply enhanced when databases and software components are well integrated. Information technologies have made drastic progresses on that way, and we could say that they have become the major key for the most mature organizations at the integration level. Even though this integration problem is still an open subject, the support of new business practices based on collaboration of partners modifies the approach of enterprise information system design. Since activities have always to be performed under the pressure of time, it is not possible to imagine an open organization without having an open information system. The ability to easily communicate between partners that are beyond the perimeter of the organization is intrinsically different from an integration view of the information system design. The ability to capture and share information seamlessly amongst information systems of different enterprises is often limited by the heterogeneity of business processes, organization units, data structures and technologies, as well as the difficulty to share knowledge about these different artefacts between partners. Collaborative information system is the term we use to call the information system that is attempted to work at the level of the network. Designing and running such a collaborative information system, keeping in mind the goal of a minimal effort for each actor to participate, is an overall issue to which we try to contribute.

This particular question receives a lot of attention that considers it as an enterprise interoperability concern. Among many definitions, interoperability is defined as the ability of two or more systems or components to exchange information and to use the information that has been exchanged [2]. New architectures and technologies have been emerging in order to deal efficiently with the interoperability problem. When the systems that have to interact are information systems of independent organizations, the definition of interoperability has to be refined with the objective to explicitly reduce the complexity of problem formulation. In the first step, we will address it at two complementary levels giving a set of hypothesis:

- at the business level, collaboration is driven by communications and messages that are supposed to be defined and controlled by specific business processes, so-called collaborative business processes (CBP). The CBP executions contribute to the achievement of common objectives in the collaboration space. They are a mean to control the coordination between the tasks performed by partners.

- at the technological level, information resources like data and information technologies like software components are assumed to be defined at the interface of each individual information system. This level is in charge of CBP executions, that is to say that the information flows are managed between software components that make the expected operations in compliance with the required structure of the CBP.

In the second step, we will add a brand new level, the organizational level, to the two previous ones. At the organizational level, in each independent organization, the actors who are involved in the collaboration are known. These actors are in charge of the activities defined in the CBP of the business level. They are also able to supervise the relevant resources and technologies at the technological level. These actors are aware of the processes in which they are involved.

Based on this framework, we propose to define a collaborative information system structure which is composed of two types of components:

- native information systems of partners who are involved in collaboration are the business added value components. Their contribution is to ensure collaboration performances. The components could behave independently from each other, and they are configured to fulfil the needs of the collaboration.

- mediation components are mainly in charge of coordinating between the business added value components that should be conformed to the specification imposed by the CBP. The parallelism or the synchronisation of activities defined in the business added value components is done by the mediation components. Moreover, such components are able to guarantee the exchange of information between the business added value components. Indeed, all types of heterogeneity between the native information systems of partners are tackled by the mediation components.

The concept of Mediation Information System (MIS) has been previously introduced by the authors in order to deal with this enterprise interoperability challenge [3]. A MIS is related to a collection of mediation components and is therefore an important part of a collaborative information system. The MIS is relevant at all the levels that we have defined before: business, technology and organization. This simple perspective about the structure of a collaborative information system satisfy the requirement of a less binding between components, i.e relative autonomy and low adaptation needs for partners, before and during collaboration. The overall system architecture that results from this choice is indeed very close to the concept of system of systems [4]. Mediation is a word that is often used in scientific works related to interoperability, either when one is explaining a federated approach while dealing with architectural aspects, or when one is focusing on data or service semantics while dealing with reducing heterogeneity barriers. Our proposal on MIS includes these two views in an emancipated approach of a solution, and in some manner, could be introduced as a business oriented extension of the middleware concept.

As the native information systems of partners are considered as predefined components that are not supposed to be modified when they participate in any collaborations according to an interoperability philosophy, most of the collaborative information system engineering effort is linked to the MIS design and operation. The MIS engineering process that we used as a main guideline is depicted in Figure 1. It follows Model Driven Engineering (MDE) principles [5].

**Insert Figure 1 here**

Classically, the model driven architecture consists of three abstraction levels: Computer Independent Model (CIM), Platform Independent Model (PIM), and Platform Specific Model (PSM). The work presented in this paper address the transition from the knowledge available before the preliminary MIS design and expressed by the partners to the result expected at the CIM level. Many different works have been performed, in complement to the one presented here, to put all the steps depicted in Figure 1 into practice. We will discuss during our presentation on some of the constraints that are imposed by the compatibility with the other steps, from CIM to PIM, and from PIM to PSM. Let us simply consider here that PIM has been defined using a Service Oriented Architecture (SOA), and PSM has been conceived with an Enterprise service Bus (ESB) as an ultimate target platform [3].

According to [5], the CIM is a model of a system that shows the system in the environment where it will operate. It helps in understanding a problem and defining a shared vocabulary for use in the models of the other levels. In our case, this CIM level concerns the organization, objectives, business, processes, and responsibilities. Thus, collaborative process model has been selected as a good candidate for representing interactions between enterprises, including the data exchanges, resources exposed to others. The CIM will be represented using the BPMN modelling formalism. So, our objective is to be able to catch, adapt and transform different kinds of knowledge about the collaboration with the aim to produce a CBP model, compliant with the requirements of the CIM level of our MIS engineering approach.

A knowledge-based system (KbS) has been developed in order to support the design of this CBP model. The system consists of three parts (Figure 2): 1) knowledge gathering to build the collaborative network model, 2) knowledge representation and reasoning to define fragments of the future CBP model, and 3) collaborative process modelling that will assemble the result obtained from the second part. The core of the KbS is the knowledge representation and reasoning part which delivers candidate fragments of the final CBP model from the collaborative network model using an ontology-based approach.

**Insert Figure 2 here**

The last part of the KbS, the collaborative process modelling, has been divided into two phases: specific knowledge extraction and BPMN construction. This leads us to obtain finally the four main functionalities as shown at the bottom of Figure 2. A prototype has been conceived, developed and implemented in order to support these four functionalities. The technical architecture of the prototype is shown below:

**Insert Figure 3 here**

The paper is structured in the same way as the KbS works. In each section, explanations will be devoted to the theoretical aspects related to functionalities of the KbS as well as the practical aspects related to the part of the prototype that implements the relevant functionalities. Section 2 will present the knowledge gathering in more details. An editor that we developed in order to support the definition of collaborative network model will be described. The major contribution is described in the section 3 including the description of the ontologies and their concepts, as well as how they are used to produce the knowledge for building the CBP model. Section 4 explains the last step, the construction of the process which is the expected result of the CIM proposal.

## 2. Knowledge gathering

The knowledge gathering focuses on capturing all necessary knowledge concerning the collaboration that is going to take place. This knowledge concerns both network characteristics (relationships between each pair of participants, and common goals), and actor profiles (roles, and services).

### 2.1.1. Inter-enterprise collaboration and knowledge

According to [7] [8] [9], we can define that collaboration has individual and collective aspects. The individual aspects concern the actors who accomplish the collaboration tasks. The collective aspect concerns the strategies, goals, relationships, as well as processes.

Collaboration leads to set up collaborative network using graphs which are generally composed of nodes (individual partners), and links (relations and flows that tends to describe collective artefacts) [10]. Several parameters for configuring collaborative networks have been defined and adapted from the enterprise modelling theory [11], such as partners, common goals describing the expectation in terms of result of network, duration, stability, relationships between partners, and organizational structure. The organizational structure concerns the *topological* definition of the network which explains how partners connect to each other through what kind of *relationships*.

The enterprise knowledge is known to be explicit, tacit, and social [12] [13] [14]. It can be acquired from many different sources in many different ways: experience, practice, conversation, innovation, document, software code, etc. The knowledge that drives and supports collaboration is for example, core competences of enterprise, experiences from previous collaborations, knowledge of interoperability issues, decision-making support, etc. [15]. However, the precision of collaboration characterisation depends largely on the knowledge we can retrieve from partners. More quantity and quality of knowledge captured leads to a more accurate characterisation of collaboration and the result will be closer to the reality.

### 2.1.2. Functionality and tool

We developed the Network Editor (NE) to support the knowledge gathering. The objective of the NE is to be used as an aided design tool in the collaborative network domain. Users of the NE are in charge of collecting the relevant knowledge by interviewing partners about their view of the collaboration space. Then, they use the NE to design the relevant collaborative network models and by the way capture and store the corresponding knowledge in a predefined pattern.

The NE should provide a design space with some tools which allow users to create, and characterise their collaborative networks in a graphic way. This editor requires some expertise, as well as effective and efficient communication between the NE's user and the network partners. We selected the Graphical Modeling Framework (GMF) for developing the NE. This tool allows us to develop complex controller logic that map a model to the different elements of a view. It tends to become a keystone framework for the rapid development of standardised Eclipse multiview graphical modelling editors. The Figure 4 shows the interface of the NE:

**Insert Figure 4 here**

- Tool palette: a set of tools allowing users to create elements on the design space. The tools that the NE offers are used for creating participant, abstract service, role, continuous and discontinuous relationships, topology and common goal. These are the essential elements that we have selected for defining a collaborative network and that will be described later on.
- Design space: an empty space for drawing collaborative network diagram by using the tools in the palette. We use this space to graphically characterize the collaborative network.
- Property sheet: a set of attributes that we have to define when creating an element. The attributes are for example name, description, type, etc. The value of each attribute can be specified by selecting from a given list (if it is enumeration) or filling in from scratch.

Thanks to the NE, we obtain a diagram file representing graphically the collaborative network and its associated XML file. This XML file is called collaborative network model which will be imported into the Knowledge Base (KB) for manipulating afterwards.

3. **Knowledge representation and reasoning**

Knowledge representation and reasoning is a fundamental aspect in the construction of knowledge-based systems. It aims to design computer systems that reason about a machine interpretable representation of the world, in a similar way to human reasoning. Reasoning refers to inferring new statements (conclusions) from a set of given ones (assumptions) which have the property that they are true whenever the assumptions are true [16].

In knowledge-based systems, there are three categories of knowledge [17]: domain, inference, and task knowledge. The domain knowledge describes the concepts, properties and instances for a particular domain. The inference knowledge is composed of inferences which are primitive reasoning steps operating over the knowledge base by inference engines. The task knowledge refers to an abstraction over the inference knowledge to promote the reuse of knowledge and knowledge-based system. The first category of knowledge captures the static knowledge, while the others define the dynamic knowledge of the system.

Ontology and rules seem to be the appropriate formalisms to handle the static and dynamic behaviours of knowledge according to [17] [18]. Ontology supports reuse of knowledge and knowledge base but lacks the expressivity for problem-solving, while rules can deal better with the problem-solving and dynamic behaviours of knowledge-based system.

An ontology-based approach has been developed for dealing with the knowledge representation and reasoning which is the core of our KbS. We defined an ontology called Collaborative Network Ontology (CNO) including the deduction rules. The CNO covers the domains of collaborative network and collaborative network process. The following sections will present the related ontologies for defining the CNO and then the CNO itself.

### 3.1. Related ontologies

We intend to use the ontology as a means for representing knowledge. [19] defined that an ontology is a formal explicit specification of a shared conceptualisation for a domain of interest. Ontology represents knowledge that can be used and reused in order to facilitate the comprehension of concepts and relations in a given domain as well as the communication between different domain actors.

In the last decade, ontologies have been developed for different purposes and cover various domains such as medicine, tourism, etc. Here we are interested in ontologies related to the business process and enterprise modelling domains. The AIAI (Artificial Intelligence Applications Institute) ontology [20] is focused particularly on intra-enterprise modelling in order to ensure a consistent communication between humans or software applications. The TOVE (TOronto Virtual Enterprise) [21] and CNO (Collaborative Networked Organization) of ECOLEAD ontologies are focused more on the virtual enterprise modelling. Both are conceptualised at the organizational level taking into account, such concepts as participant, role, and activity. The Business Process Management Ontology (BPMO) [22], Process Specification Language Ontology (PSL) [23], and MIT Process Handbook Ontology (PH) [24] [25] are oriented to the process modelling at the functional level. The BPMO provides a stable platform for the definition of business processes in order to better align information technology with business. The PSL ontology was originally created to represent processes for the specific domain of manufacturing applications. The PH ontology is more generic and can be applicable to any industries and business domains.

Since we intend to develop an ontology that covers the inter-enterprise collaboration and collaborative business process domains, the nearest to our intention is the PH ontology. Besides, we can reuse the business process knowledge from the PH to define collaborative processes. We present our ontology including its concepts, and relations between them in detail in the next section.

**3.2. Collaborative Network Ontology (CNO)**

We developed our CNO on the basis of the characteristics of collaboration that are coming from the knowledge gathering and of the characteristics of collaborative process that are based on knowledge reuse from existing ontologies [26]. The CNO consists of ontologies (concepts, relations and properties) and also rules for dealing with the dynamic and static behaviours of knowledge. The ontologies in the CNO are: Collaboration Ontology (CO) and Collaborative Process Ontology (CPO) representing the elements of collaborative network and process respectively. The deduction rules connect these two ontologies together by the semantics and structural links which reflect the notion of consequence and make deductive reasoning possible in the KbS. Moreover, the deduction rules are really important in our knowledge-based system because it ensures the morphing of the collaboration knowledge (in CO) into the collaborative process knowledge (in CPO).

The most common ontology language for editing ontologies is OWL (Web Ontology Language). We selected OWL-DL sublanguage because we intend to carry out automated reasoning and we may need to define more than one cardinality for some concepts in our ontology.

**3.2.1. Collaboration Ontology (CO)**

The CO refers to the conceptualisation of enterprise collaboration and is mainly based on the inter-enterprise collaboration model as discussed in the knowledge gathering part. The CO schema is shown in the Figure 5:

**Insert Figure 5 here**

- Collaborative network is a group of at least two participants who work together in response to one or multiple common goals and a set of relationships between the participants.
- Participant can be a physical actor or an enterprise that joins the network in order to carry out a common goal collaboratively with other participants.
- Role defines the responsibility of participant in the network. For example: seller, buyer, producer. It refers to the resource concept of the PH ontology.
- Abstract service is a high-level service that explains the competencies or the know-how of the participant. For example: marketing and sale, procurement. This concept comes from the business activity model (BAM) approach of the PH ontology [24].

- **Common goal** describes the reason why the network is established in terms of products or services to deliver to customers [11]. It gives the direction the partners have to head for and achieve.
- **Relationship** defines the interaction between two participants. It describes how partners connect to each other. Three types of relationship have been classified [27]: competition (if enterprises are in the same sector of business), supplier-customer (if enterprises collaborate with their partners who supply them complementary services), and group of interest (if enterprises are neither in substitutability nor vital complementary, but annexed additivity).
- **Topology** describes the overall relationship structure of the network. Three basic forms of topology are defined based on the circulation flow [28]: chain, star, and peer-to-peer. The form of topology can be distinguished by the orientation of decision-making power and duration of collaboration in the network.
- **Decision-making power** describes the orientation of decision-making in the network. Three types are distinguished: central, equal or hierarchic.
- **Duration** describes the frequency of interactions that occur during the collaboration. [11] distinguished two kinds of duration: continuous or discontinuous.

### 3.2.2. Collaborative Process Ontology (CPO)

The CPO refers to the conceptualisation of collaborative process. The CPO is an extension of the PH ontology [25] and the Meta-Model of Business Collaborative Process (MMBCP) introduced in [3]. The CPO covers the concepts of business service, resource, dependency, coordination service, and MIS service. Only the MIS service concept comes from the MMBCP, while the others are inspired from the PH ontology. In fact, the dependency concept of the PH ontology can be considered as the message and sequence flows of the MMBCP. The coordination service concept is the main point for connecting the PH ontology to the MIS service concept of the MMBCP. The CPO schema is shown in the Figure 6:

**Insert Figure 6 here**

- **Business service** explains a task at functional level. An abstract service is composed of some business services. For example: obtain order, receive products. This concept is inspired from the BAM approach of the PH ontology [24].
- **Resource** can be data, machine, tool or material used or produced by business service. Resource concerns the resource concept defined in the PH ontology.

- Coordination service is used for managing the resource dependency. For example: manage flow of material, manage accessibility of documents. This concept comes from the model of collaborative process concept of the PH ontology [24].
- MIS service is defined in the MMBCP [3]. We consider a coordination service as a MIS service because both are collaborative service provided by the collaborative platform (or MIS).
- Dependency between business services (message flow) is a flow from a business service to another when they have a resource in common. The two business services linked by this kind of flow do not belong to the same participant.
- Dependency between MIS services (sequence flow) is a flow between two MIS services when they have a resource in common.

### 3.2.3. Deduction Rules

We defined the deduction rules based on our expertise and some references found in the literature. We specify five groups of rules (GRs). Each group may have several rules defined which all work basing on the same concept. We show here only one rule for each group with an example of deduction that can be done.

The rules are written in Semantic Web Rule Language (SWRL) [29]. It is designed to be used in the context OWL-DL and thereby inherit important semantic characteristics that make automated reasoning more tractable.

#### 3.2.3.1. GR.1: Role and abstract service

This group is intended to derive abstract service and role when each is provided. According to [30], each virtual enterprise is represented by its goals, the activities to achieve the goals, the roles that perform the activities and the skills that are required to fill the roles. This definition refers to the relation between role and activity. We consider this activity as abstract service which describes competence of its provider. So, we defined two rules in this group. This is a rule that derives abstract service when role is recognised:

*Participant(?x)* ∧ *playRole(?x, ?y)* ∧ *performAService(?y, ?z)* → *provideAService(?x, ?z)*

This rule starts at retrieving roles of participant and finding abstract services that can be performed by these roles. Then, the rule will return the list of abstract services that correspond to the roles the participant plays. The Figure 7 illustrates how the above rule works by showing instances with respect to their corresponding concepts in the ontology:

**Insert Figure 7 here**

From above, the instances are represented as ellipses. The instance linked with dash-dot line is the one we are defining. Those linked with dotted lines are what the rule derives. Those linked with dashed lines already existed in the Knowledge Base (KB) before running this rule. Otherwise, the rule does not function as it should. The figure explains that if the network has participant *A* who plays the role of *seller*, then participant *A* provides the *sell service*, *sell product*, and *sell items from stock* abstract services.

Another rule in this group works vice-versa. This means the rule derives role when abstract service is given.

### 3.2.3.2. GR.2: Business service

This group concerns the deduction of business services when an abstract service is recognised. We define this rule on the basis of the BAM of the PH [24] which states that every abstract activity has its corresponding functional-level activities. We consider these functional activities as business services. There is only one rule in this group which is written as follows:

*Participant(?x)* ∧ *provideAService(?x, ?y)* ∧ *hasBusinessService(?y, ?a)* → *provideBusinessService(?x, ?a)*

This rule starts at retrieving business services that correspond to the abstract services provided by the participant. Then, the rule will return the list of business services that the participant should expose. This rule is the key that creates the semantic connections between the CO and CPO via the concepts of abstract and business services.

From the example of the first rule, for the *sell product* abstract service of *A*, *A* has to provide the *obtain order*, *prepare products to deliver*, and *transfer invoice* business services.

### 3.2.3.3. GR.3: Dependency and MIS service

This third group concerns the deductions of dependencies for both message and process logic sequence flows, coordination services, and MIS services. The rules defined in this group are the most complicated ones because they take into account several concepts at the same time.

Here below is an example of the rules of this group. This rule allows us to deduce dependencies when two business services belonging to different participants have a resource in common. The resource dependency concept has been defined by [24]. Coordination services can be deduced from dependencies by taking into consideration the exploitation of resources. [31] affirms the relation between dependency and coordination service, whereas [2] supported the idea that the coordination service can be considered as the MIS service.

*CNetwork(?a)* ∧ *hasRelationship(?a, ?z)* ∧ *P1(?z, ?y)* ∧ *provideBusinessService(?y, ?c)* ∧ *hasOutput(?c, ?d)* ∧ *P2(?z, ?x)* ∧ *provideBusinessService(?x, ?b)* ∧ *hasInput(?b, ?d)* ∧ *CoordinationService(?f)* ∧ *manipulateResource(?f, ?d)* ∧ *Dependency_between_BusinessServices(?e)* → *fromBusinessService(?e, ?c)* ∧ *toBusinessService(?e, ?b)* ∧ *containResource(?e, ?d)* ∧ *isCoordinatedBy(?e, ?f)* ∧ *hasMISservice(?a, ?f)* ∧ *MISservice(?f)*

This rule starts by finding a relationship between two participants via P1 and P2 relations. Each participant provides its own business services which have input and output resources. The rule verifies whether the output of a business service is the same as the input of other business services. If so, the rule finds a coordination service that can manipulates such resource and creates dependency between these two business services. It also defines this coordination service as a MIS service.

To continue the example of the previous rules, we suppose here that there is another participant *B* in the network and assume that the participants *A* and *B* establish a relationship between them. From the second rule, we obtained that the participants *A* and *B* provide *obtain order* and *place order* business services respectively. The *place order* service has *purchase order* resource as output, while the *obtain order* service has the same resource as input. The current rule deduces a dependency of *purchase order* between these two business services. The *manage flow of document* coordination service can manipulate the *purchase order* resource. This coordination service is also created as the MIS service.

Since the rule is described by considering the direction of resource flows between business services, for each service we have to consider both its inputs and outputs. The rule shown earlier is presented by the *dependency 1* of the Figure 8:

**Insert Figure 8 here**

Therefore, there is another rule dealing with the direction of the *dependency 2*. The dependency between business services belonging to different participants is called message flow. Not only these two rules that take care of the dependencies 1 and 2, there is another important rule in this group that concerns the dependency between MIS services (sequence flow). Such rule is also based on the same input-output concept as described above.

### 3.2.3.4. GR.4: Common goal

This group is dedicated to deducing a list of abstract services to be included in the network. The abstract services deduced by the first rule are the ones that the involved partners provide to the others. They are a subset of the abstract services obtained by this actual rule. There is only one rule in this group which derives abstract services from goal:

*CommonGoal(?x)* ∧ *description(?x, ?a)* ∧ *swrlb:substringBefore(?y, ?a, " ")* ∧ *AbstractService(?b)* ∧ *name(?b, ?c)* ∧ *swrlb:containsIgnoreCase(?c, ?y)* → *achievesAService(?x, ?b)*

The rule starts by segmenting the description of common goal and keeping only the first word found. Then the rule searches in the KB for abstract services whose name contains this word. The abstract services obtained are the services that concern all involved partners and the network itself have to provide.

We adopt the concept of goal from [32], which defines a goal consisting of verb and parameters (profit, direction, result). However, the limitation of SWRL built-ins (prefix *swrlb*) does not allow us to analyse goal as discussed in the original concept because some built-ins have not yet been implemented. Consequently the rule has not yet been completed. Furthermore, there are restrictions in term of expressing description of a common goal since it is required to start with a verb. The full implementation should take the whole phase of description into account and analyse every segment of it.

For example, for the *buy 100 bolts* common goal, the rule deduces *buy*, *buy over internet*, and *buy in a store* abstract services. All of these abstract services contain the *buy* which is the first word of the description of common goal.

### 3.2.3.5. GR.5: Topology

The rules in this group are dedicated to deducing the type of topology when the orientation of decision-making power and the duration of communications are provided. This group has three rules which are all shown below:

*Topology(?x)* ∧ *hasPower(?x, central)* ∧ *hasDuration(?x, continuous)* → *hasType(?x, star)*

*Topology(?x)* ∧ *hasPower(?x, equal)* ∧ *hasDuration(?x, discontinuous)* → *hasType(?x, P2P)*

*Topology(?x)* ∧ *hasPower(?x, hierarchic)* ∧ *hasDuration(?x, continuous)* → *hasType(?x, chain)*

These rules are specified on the basis of the characteristics of topology (chain, star, and P2P) [28]. The way we describe the rules in this group is different from the others. We specify the instances of concepts directly in the SWRL rules. If the concepts meet the instances defined, the rules return the instance result. For example, topology is *star* if it has *central* power and *continuous* duration.

However, these rules can be defined and reasoned directly by the ontology, but we decided to define them as separated rules just because of the compatibility reason with the other rules. Furthermore, we can distinguish them easily as a group of rules not the restrictions defined in the class.

### 3.2.4. Tool

To constitute the Knowledge Base (KB), we use the CNO (CO, CPO and deduction rules) defined earlier. It is an OWL-based ontology. We developed the KB with the Protégé [33] tool. One of the most important aspects of a knowledge base is the quality of instances it contains. The instances we store originally in our KB come from the dataset [34] which is an OWLized version of the PH ontology because the CPO of the CNO is mostly based on the PH ontology. The dataset provides approximately 5000 instances of processes, goals, resources including roles. These instances are generic and can be used to define many kinds of processes. The Figures 9 and 10 illustrate some instances of the class *Abstract Service* and SWRL rules stored in the KB:

**Insert Figures 9 and 10 here**

Our KB takes as input the collaborative network model obtained from the NE because this network model describes the characteristics of collaboration (partners, relationships, topology, common goal, role and abstract service) which corresponds to the concepts and relations defined in the CO. Since this network model of the NE is an XML-based model but the KB requires an OWL-based model, a model transformation is needed at this step. This transformation deals with XML-to-XML transformation since OWL is also based on XML [35]. XSLT (XML Transformation) appears as the most outstanding XML model transformation language [36]. Since models can be serialized as XML using the XMI (XML Metadata Interchange) implementing model transformation, using XSLT seems very attractive.

This transformation is quite simple and direct because the names of the elements of the collaborative network model are mostly the same as the ones of the CO's elements. The concepts and an example of such transformation are shown respectively in the Table 1 and Figure 11:

**Insert Table 1 and Figure 11 here**

Once the transformation has been done, we can import the collaborative network model of the NE into the KB. The KB reasons with the instances in order to deduce knowledge about collaborative process, such as business services, resource dependencies, and collaborative services.

4. **Collaborative process modelling**

The collaborative process modelling is the last part of our KbS. It concerns the extraction of knowledge and the representation of it in form of collaborative process conforming to the BPMN syntax.

Firstly, we introduce the definitions of collaborative process and why we decided to use BPMN as collaborative process modelling language. Then, we describe the different functionalities of this part together with several technologies and tools that we implemented to support them.

4.1.1. **Collaborative process definition and formalism**

Many definitions of collaborative process have been proposed in the literature. [37] pointed out that the aspect of multi-organizations is essential in a collaborative process because the partners have their specific competencies, so they provide the activities they can perform in order to achieve the objective of process. We considered the activities provided by partners as their internal process. But, we also stated that some of these activities are added value activities for the collaboration, and have to be reachable through an interface with the MIS. We can extract some interesting characteristics of a collaborative process from these definitions as follows:

- Taking place between multiple independent entities (enterprises, organizations, or individuals).
- Common objectives to be achieved.
- Implying governance between the involved entities.
- Different entities providing a specific competency and playing a specific role.
- Independent entities exchanging resources and collaboratively performing their activities to pursue the objective of the process.

In order to formally representing this CBP, [3] also introduced the MMBCP assuming a Service Oriented Architecture (SOA) context where activities are considered as services that partners expose. The MMBCP shall be viewed as a specification of language construct that is superposed to the BPMN native language definitions. In some way, it could be seen as a special BPMN profile specially adapted to CBP models. The MMBCP has been defined by referencing the BPMN specification as well as our collaboration aspect [38].

There are many languages for business process modelling, for example, flow chart, Petri net, IDEF0, PCD (Process Chain Diagram) of ARIS, activity diagram of UML (Unified Modelling Language), and BPMN (Business Process Management Notation). Some of these languages are basic languages that take into account only an activity based representation of facts. Process modelling is therefore reduced to just a sequence of activities linked either by flows of entities and/or by events that fix the execution logic. Other languages extend those capacities to include communication (data) and organizational (actors) concepts while keeping activities as a central point. This is the case for BPMN where pool and lanes provide a relationship between activities and actors, and where a typology of arrows allows to make the difference between a time constraint and a message flow. The BPMN language has been originally designed for workflow management trying to cover all the particularities of business processes that are specified for this objective. Our MIS approach is very close to that kind of needs as the CBP appears as a specification of exchanges between the business added value components emphasizing the interactions between people, activities and information flows keeping in mind their coordination with respect to time using event triggering.

One of the main constraints imposed by the MMBCP is to define a lane for each mediation component where only the relevant coordination services are allowed to be defined. This evolution has been very easily done coming to a proof that BPMN satisfies our requirements in terms of representation.

Moreover, such a choice is enforced by the availability of modelling tools that delivers XML-based files to support importation and exportation of models, which is one of the prime interests considering our MDE approach. And last but not least, the language is recommended by the OMG for such kind of applications.

4.1.2. Functionalities and tools

The collaborative process modelling part requires the implementation of several functionalities and tools. This part starts by extracting knowledge and then modelling it in fragment of the CBP model that conforms to the MMBCP [3]. The followings describe these functionalities and tools:

- **Extraction of knowledge:** The use of SPARQL (Simple Protocol And RDF Query Language) [39] is focused on querying the knowledge corresponding to the collaborative network that we are studying. Such knowledge concerns common goals, relationships, topologies with their type, participants, roles, provided abstract and business services with associated input and output resources, dependencies and their manipulated resources, and finally MIS services that coordinate those dependencies. The Figures 12 and 13 show an example of SPARQL query and its result in XML format:

**Insert Figures 12 and 13 here**

This example shows a SPARQL query for extracting the name and roles of the participants in a network. The query returns two results of *name* and *role*. This means that there are two participants in this studied network. The first participant has the *name A* and the *role seller*. Another participant has the *name B* and the *role buyer*. The XML result from the SPARQL query will be used for XSLT transformation directly in order to generate a corresponding collaborative process model for the CPE in the next functionality.

- **Collaborative process modelling and complementary concepts:** We developed an editor called Collaborative Process Editor (CPE) to support this functionality. Such editor allows users to: visualise knowledge extracted previously by SPARQL queries and work on (verify and complete) this collaborative process. The CPE has been developed with the GMF as same as the NE. The MMBCP is integrated into the development of the CPE in order to guarantee the compliance of CBP model generated at the end of our prototype with the requirements of the PIM level (Figure 1). Thus, the CBP models obtained from this functionality do totally conform to the MMBCP defined in [3] in order to have the appropriate BPMN process models at the end of the next functionality. The Figure 14 shows the interface of the CPE:

**Insert Figure 14 here**

The CPE offers a design space, a tool palette and a property sheet. But, the tool palette of the CPE contains more tools than the NE's. The additional tools are for creating: business service, resource, MIS, MIS service, gateway, event, and dependency. When we create a new collaborative process, we get a diagram file that represents graphically the collaborative process, as well as its associated XML file.

However, at this stage the collaborative process obtained is not complete yet. It does not have gateway and event which are the important modelling elements of BPMN. Thus, the generations of gateway and event from this collaborative process are necessary.

The generation of gateways is based on the patterns of dependency. Dependencies are referred to the resource flows between services. Gateways control how flow interacts as they converge and diverge within a Process [38]. We distinguished two patterns as shown in the Figure 15:

**Insert Figure 15 here**

The two patterns of dependencies concern dependencies between services, gateways, or combination of both (represented by rectangle figure). The polygon figure represents only gateway that controls multiple flows in or out.

The generation of events is based on these following rules: 1) If MIS service has no outgoing sequence flow (flow in the MIS pool), then generate a new sequence flow out from this MIS service to the end event, and 2) If business service of partners has no incoming message flow (flow between MIS and partners' pools), then generate a new message flow out from this business service to the start event.

The transformation rules have been defined with XSLT based on the above patterns and rules to automatically generate gateways and events. Nevertheless, we can only generate the figure of gateways and events, but not its type. The types will be manually specified because they need to take into account the meanings of resources in each flow as well as the collaborative process. Once these generations have been done, the partners have to validate and agree on using this complete collaborative process for supporting their collaboration. The collaborative process model obtained at the end of this functionality will be used to build a relevant BPMN collaborative process in the next functionality.

- **BPMN construction:** The transformation of the agreed collaborative process models from the previous functionality into a BPMN relevant one is done by the ATL (Atlas Transformation Language) [40]. This concerns XML-to-BPMN transformation. However, using XSLT for transforming an XML into a BPMN seems to be more complicated than the XML transformation, and requires much effort due to the complexity of BPMN meta-model. Transformation by using meta-model concepts on a very high level appears to be more realistic. Such transformation requires a mapping definition between elements of meta-models. A transformation with ATL requires a source model, a source meta-model, a target meta-model, and an ATL file. In our case, the source meta-model is the XML meta-model since the source model is an XML-based collaborative process model generated by the CPE, and the source meta-model is the XML meta-model. The target meta-model is of course the BPMN meta-model. Six rules have been defined to accomplish such transformation, as shown in the Table 2 and Figure 16:

**Insert Table 2 and Figure 16 here**

Once the transformation has been done, we visualise a BPMN process with the STP BPMN Modeler. This modeler is provided under the Eclipse Public License (EPL) and is also a GMF-based. The STP BPMN Modeler is a graphical editor specifically dedicated to BPMN. It offers the same design space, and property sheet as both the NE and CPE. But, the creation tools contained in the tool palette are dedicated to the design of BPMN diagram according to the BPMN specification of the OMG (Object Management Group). The tools are for example: pool, lane, task, message and sequence flows, and several types of gateway and event. In the same way as NE and CPE, we obtain a BPMN diagram file that represents graphically BPMN collaborative process and its associated XML file. Figure 17 shows the interface of the STP BPMN Modeler:

**Insert Figure 17 here**

From the above figure, we can see that the way the elements are arranged on this BPMN diagram corresponds to the restrictions specified in the MMBCP. We have mentioned earlier that the BPMN diagrams obtained at the last step of the prototype conform to the MMBCP as we take into account this meta-model in the CPE (previous functionality).

5. **Conclusion**

The objective of this paper is to present our approach for developing a knowledge-based system dedicated to specifying a collaborative process model specific to a given collaboration case. Our approach is based on ontologies and deduction rules to deal with the knowledge morphing from a collaboration space into a collaborative process. The knowledge-based system we developed is a prototype which is composed of three main parts: knowledge gathering, knowledge representation and reasoning, and collaborative process modelling. Some tools and open source technologies have been implemented in order to fulfil the development of this system.

However, on a practical side, if the prototype of the system works correctly, it still has several limitations that should be improved. The prototype should be more user-friendly in terms of user interfaces, design concept, and functionalities.. It is an interesting way of progress to think about a CBP editor that could be used simultaneously, in a wiki mode, by the partners in order to design collectively the specifications of their collaborative information system. We provide the background to continue works in such a perspective.

Some perspectives could also be discussed on more theoretical aspects of our study, as follows:

Firstly, the concepts, relations, and restrictions of the current CNO as well as the deduction rules can be enriched. The enrichment of CNO and rules may constrain more on the deduction and make the result of deduction more accessible to users. It is the fact that the more the process repository is rich, the more the selection of candidate fragments for the CBP model design is complicated. This part could be improved working both on the level of the knowledge gathering searching means to further discriminate candidates, and on the level of rules in order to refine the search of candidate model fragments by making more intensive correlations between them. Secondly, the generation CBP model in BPMN is not fully completed as the gateways and events are not all included in the knowledge representation and reasoning phases. Up to now, we can generate only the start and event events and four types of gateway (parallel, event-based exclusive, data-based exclusive, and data-based inclusive gateways) which are frequently used in BPMN processes. We have not included yet for example, message event, time event, complex gateway, etc. Thus, we may need to enrich this generation concept by taking into account these missing elements. This part is an on-going work that we should think about using knowledge about the control structure of the CBP.

Finally, the instances originally stored in the KB come only from the PH ontology which limits the models of collaborative process generated from our system. Thus, we need to enlarge the KB by storing more instances coming from other sources, such as an example: collaboration use cases or creating new business services by partners from scratch.

The main characteristics of our contribution can be summarised in three points:

- the approach is based on a cross fertilisation of two kinds of knowledge, one coming from data collection from the partners about their capabilities and ambitions with respect to the future collaboration space, and another coming from knowledge stored in process repositories making explicit description of individual and common activities becoming possible. This idea of splitting the whole set of knowledge into two separate ontology is a basis for a broad range of deduction possibilities, using rules that mix the knowledge elements in many different ways to select possible candidate fragments for the CBP model design. As far as we know, many studies have worked separately on each part, but little has been done using an overall approach of the problem. We have just made a proof of concept with this idea, and many perspectives could be drawn starting from this point, as explained before.

- a CBP model is built on the basis of a meta-model that meets the requirements for a model driven engineering process of a mediation information system that will support the collaboration in runtime. The CBP model is not a final result, but appears as in intermediate artefact inserted in a more global design process. As a consequence, the integration of our tools to the other set of MDE tools is powerful. It offers new capabilities such as to define a kind of agility in the design of the system which will be helpful if a change in some requirements should be taken into account during design time or runtime. The definition of activities in the collaboration ontology and in the collaborative process ontology is done with the service concept in the knowledge structure. So, the deduction rules give an opportunity to address the mapping between abstract services and business services in a very flexible manner. The context in which each type of service is identified is particularly rich, and by consequence, many different possibilities to draw links between them are included in our problem formulation and solving.

**Figure Captions**

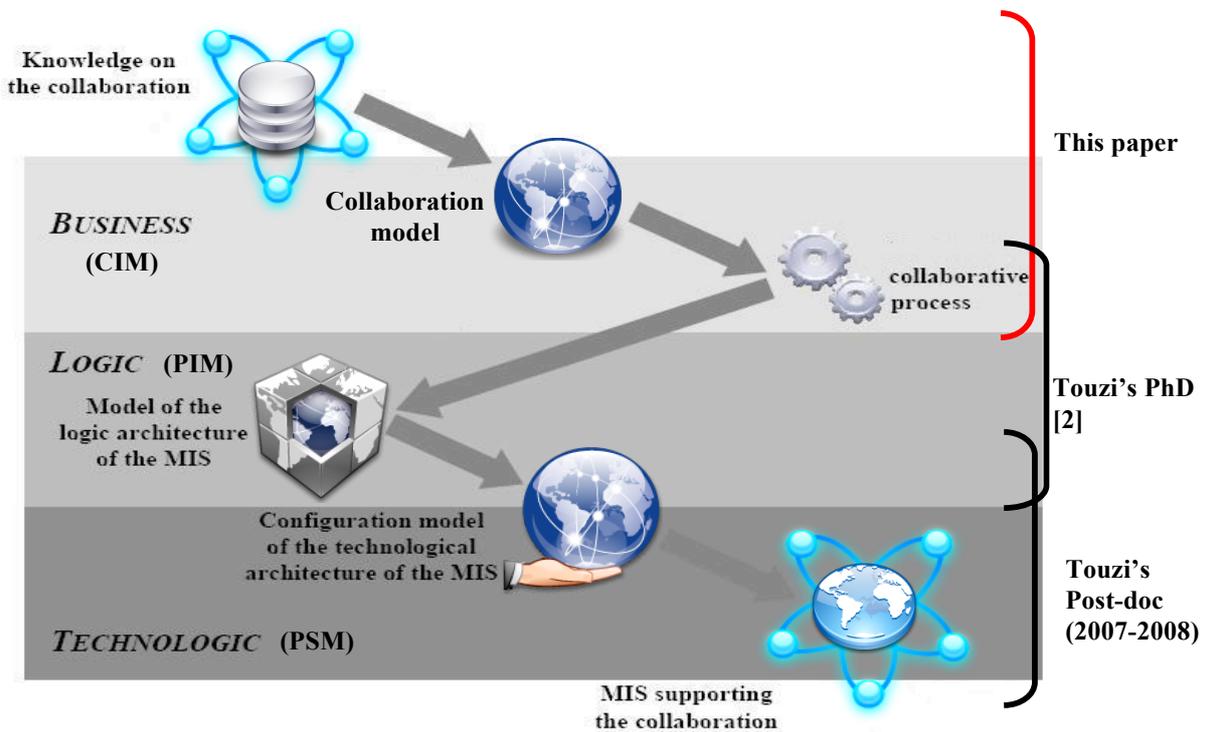

Figure 1 Global principle of MIS design through a Model Driven Approach [6]

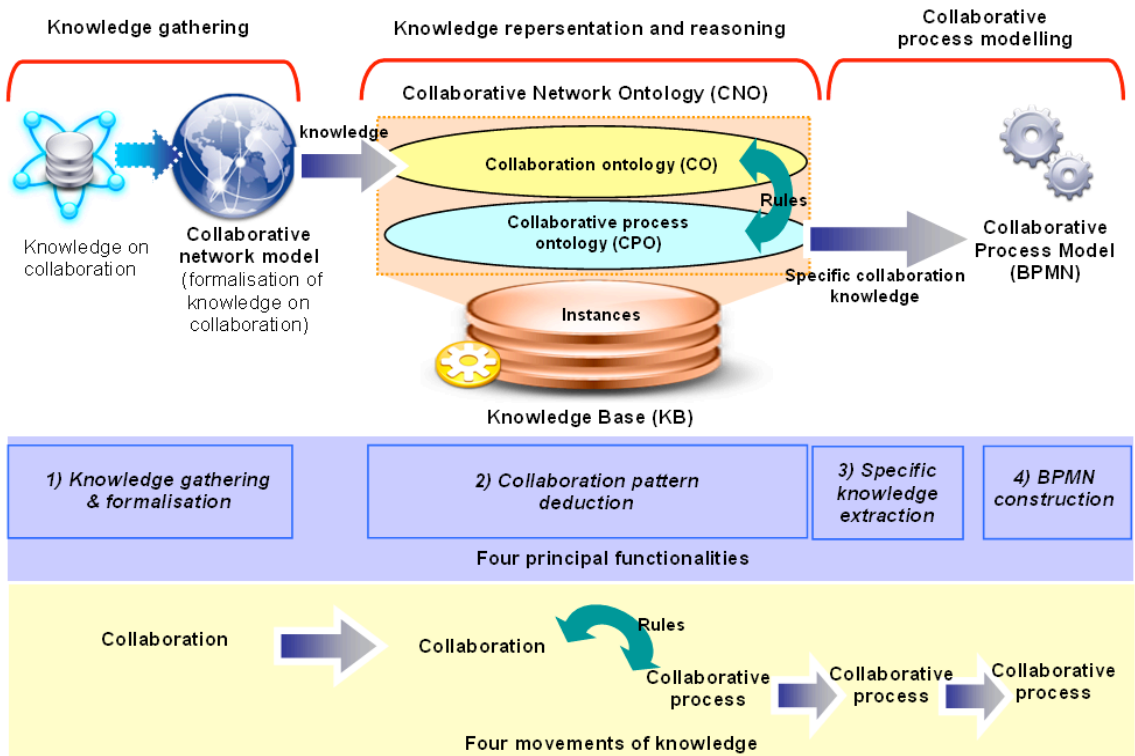

Figure 1 Knowledge-based system (KbS) and functionalities

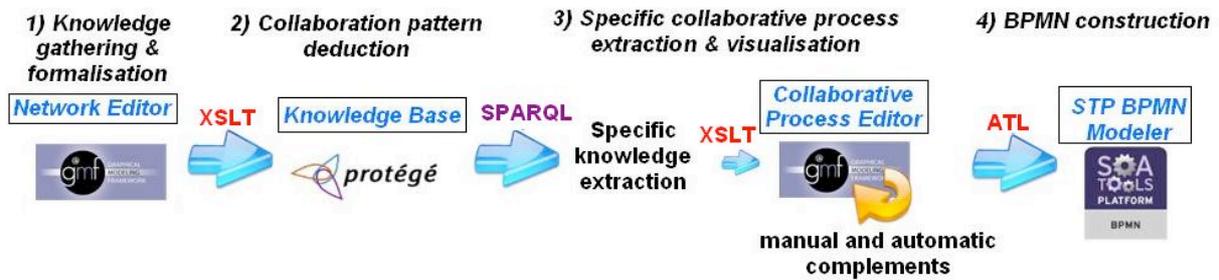

Figure 3 Four functionalities of the prototype and development technologies

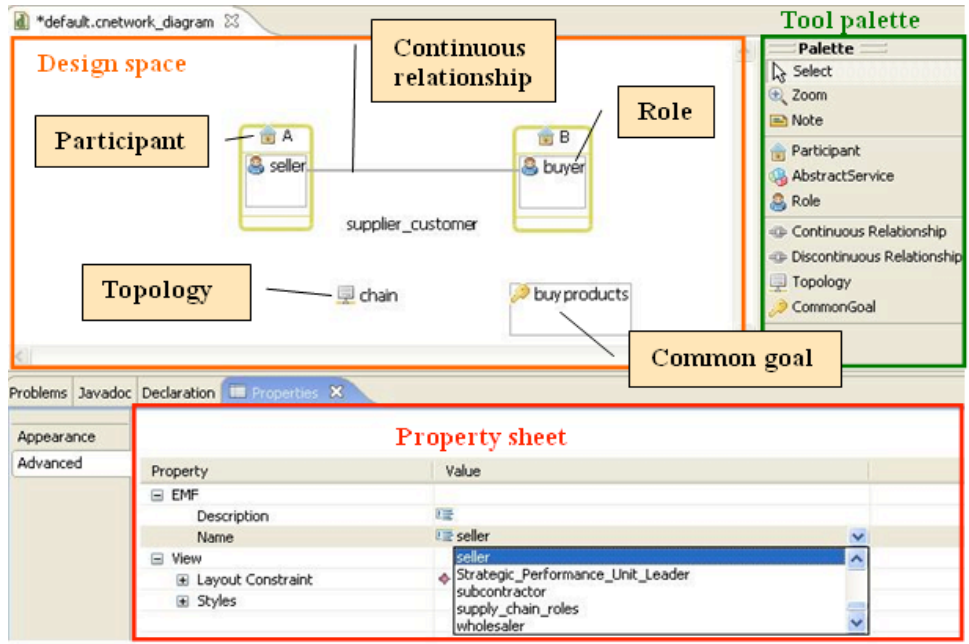

Figure 4 Interface of the NE tool

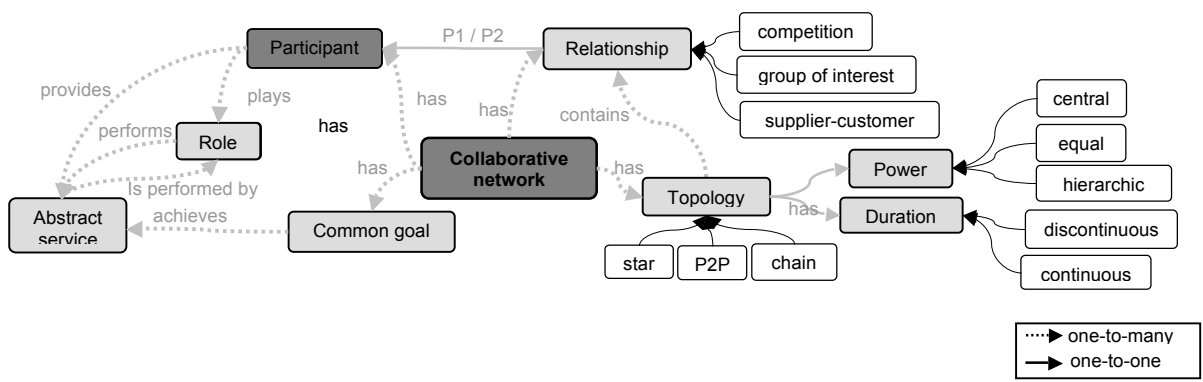

Figure 5 Collaboration Ontology (CO) schema

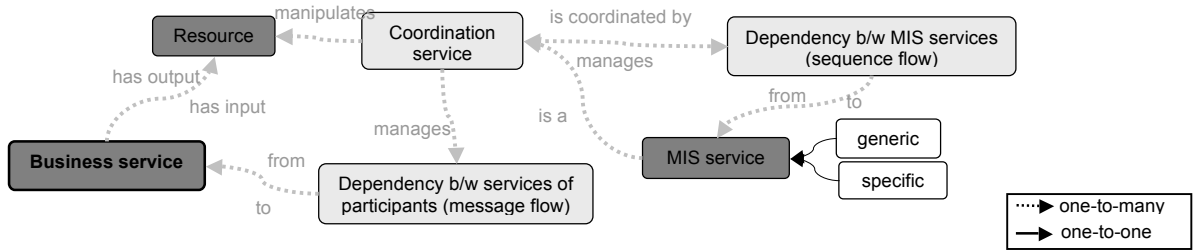

**Figure 6 Collaboration Process Ontology (CPO) schema**

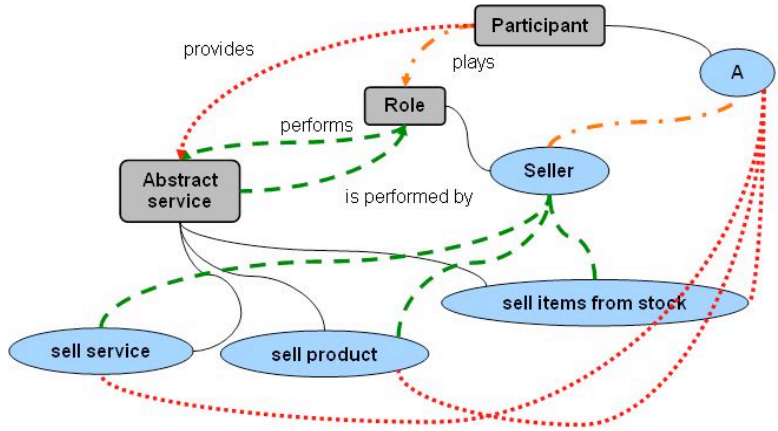

**Figure 7 Example of deduction by the GR.1**

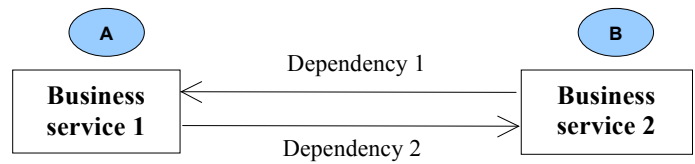

**Figure 8 Two-way dependency consideration**

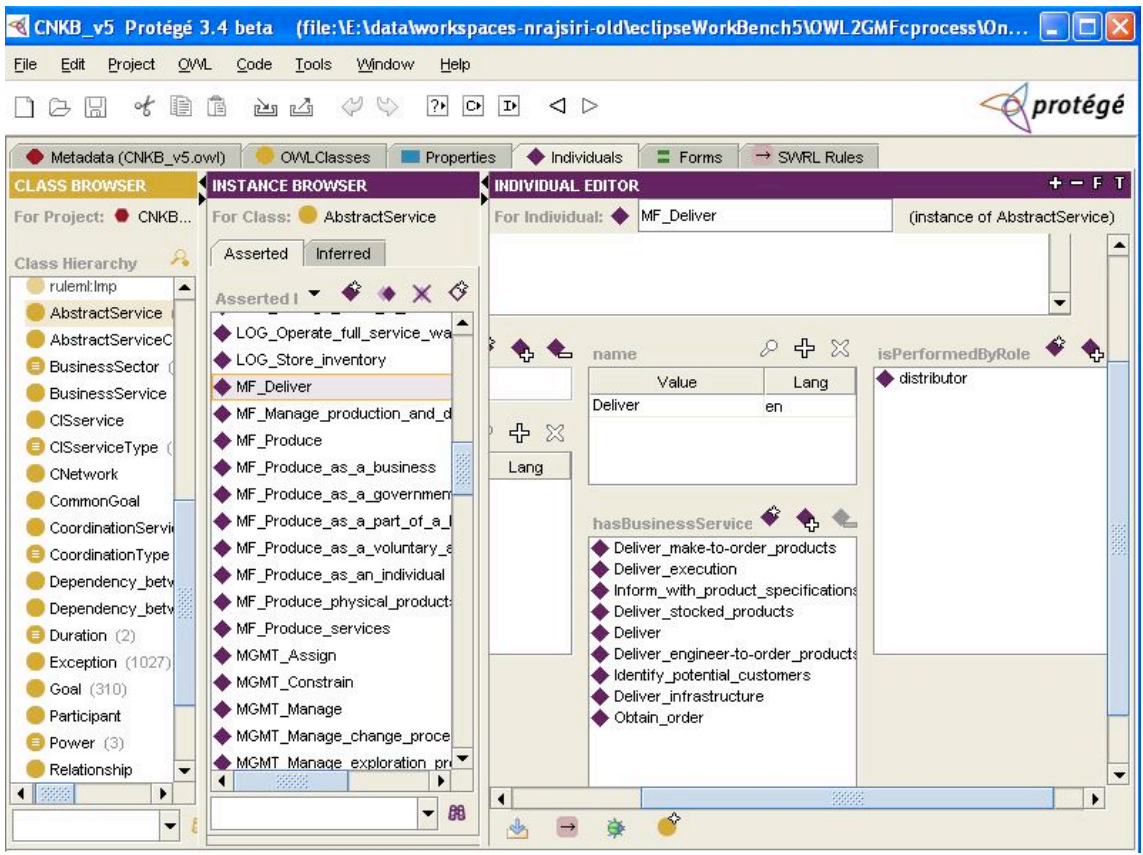

**Figure 9 Instances of the class Abstract service**

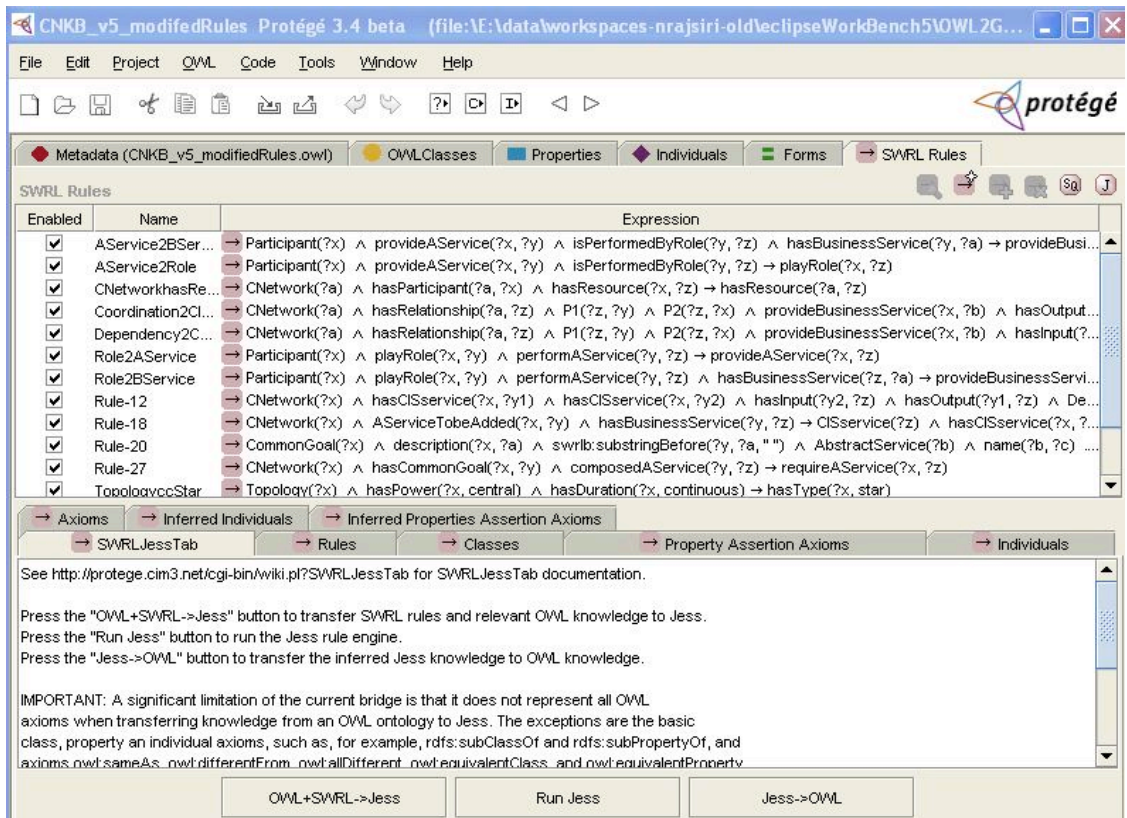

**Figure 10 Instances of the class Abstract service**

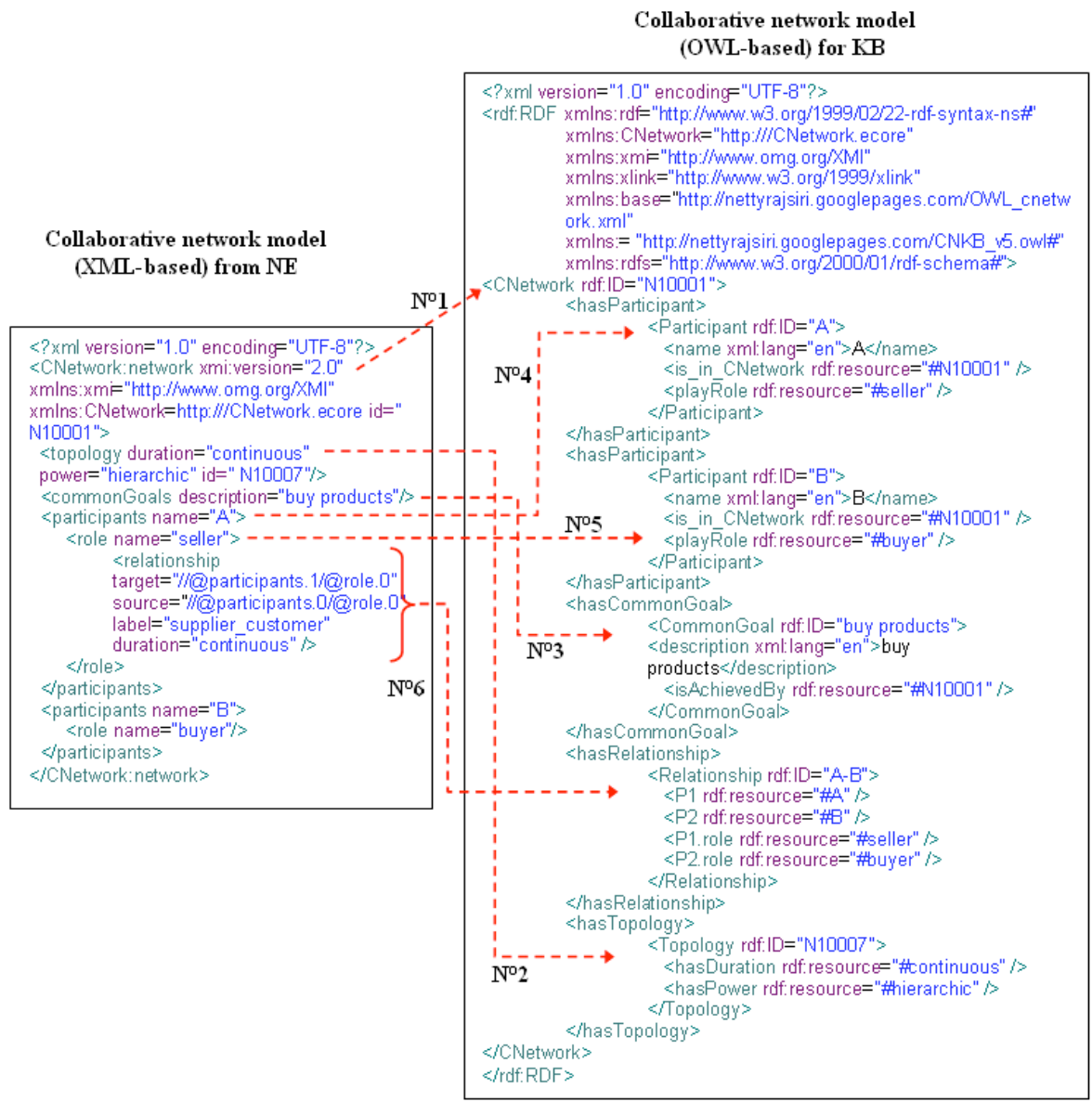

**Figure 11 Transformation with XSLT of a collaborative network model (XML-based) from NE into a model (OWL-based) for KB**

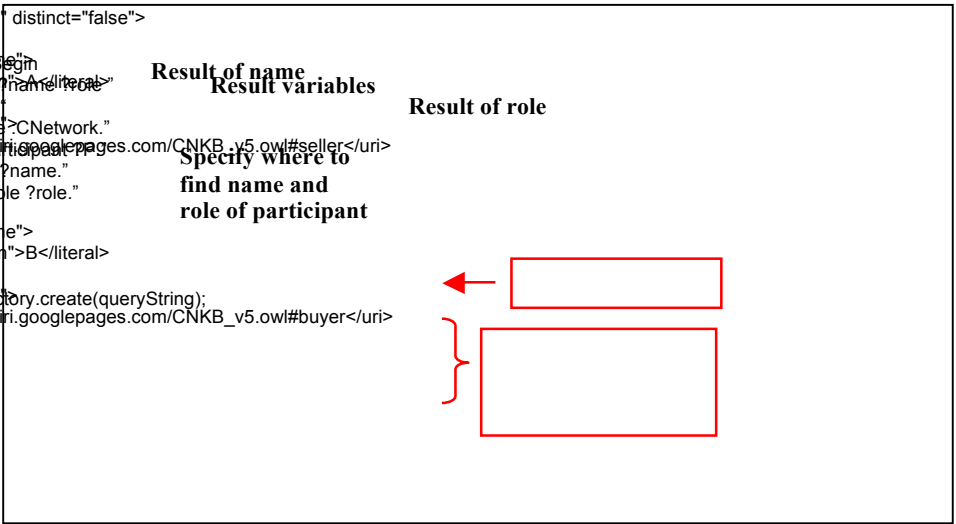

Figure 12 SPARQL query to extract name and role of the participants in a network

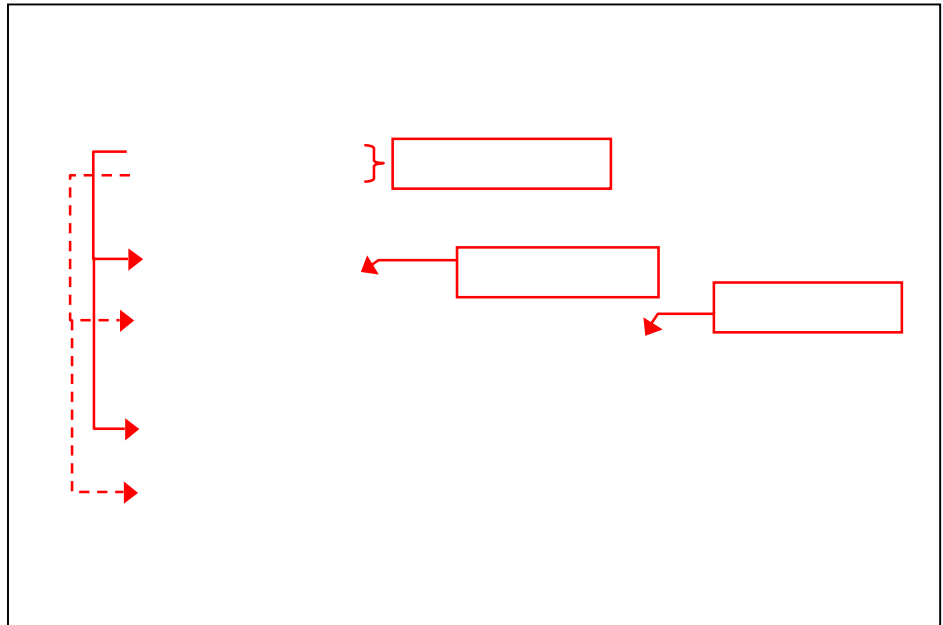

Figure 13 SPARQL query result in XML

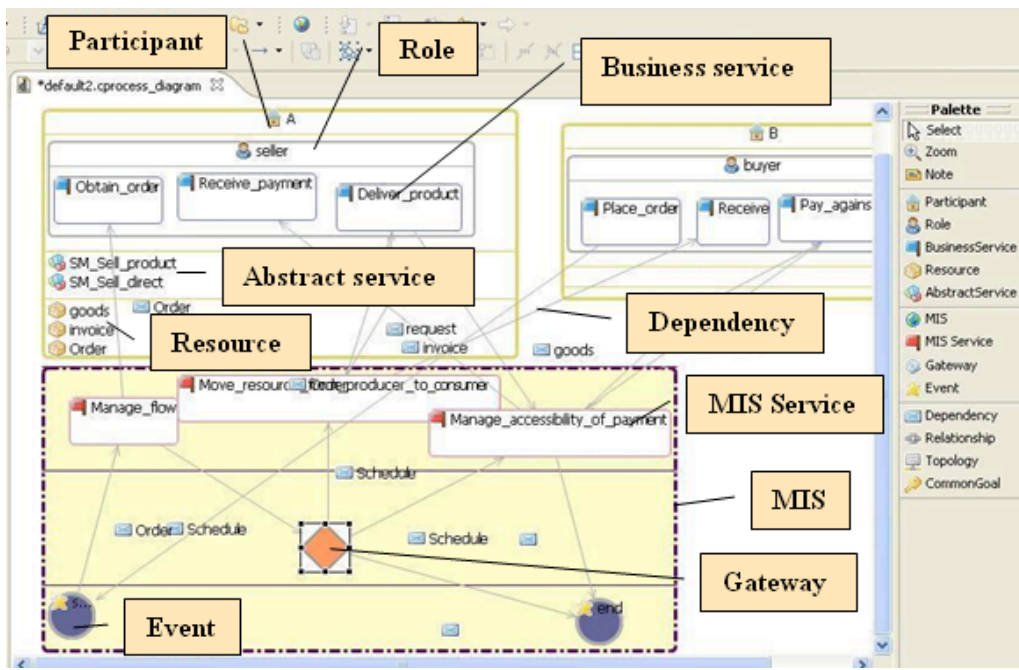

Figure 14 Interface of the CPE tool

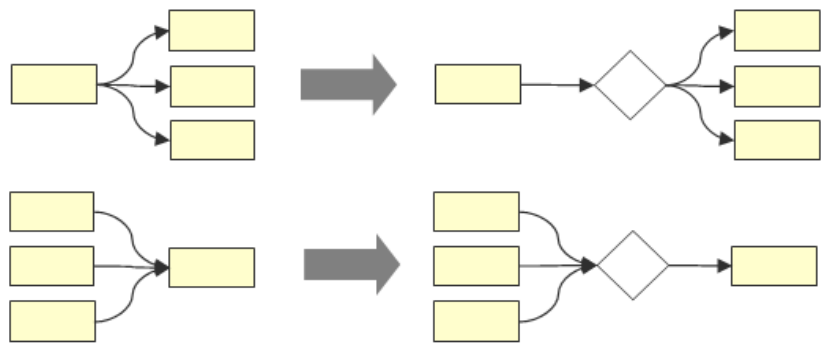

Figure 15 Patterns of dependencies and gateway

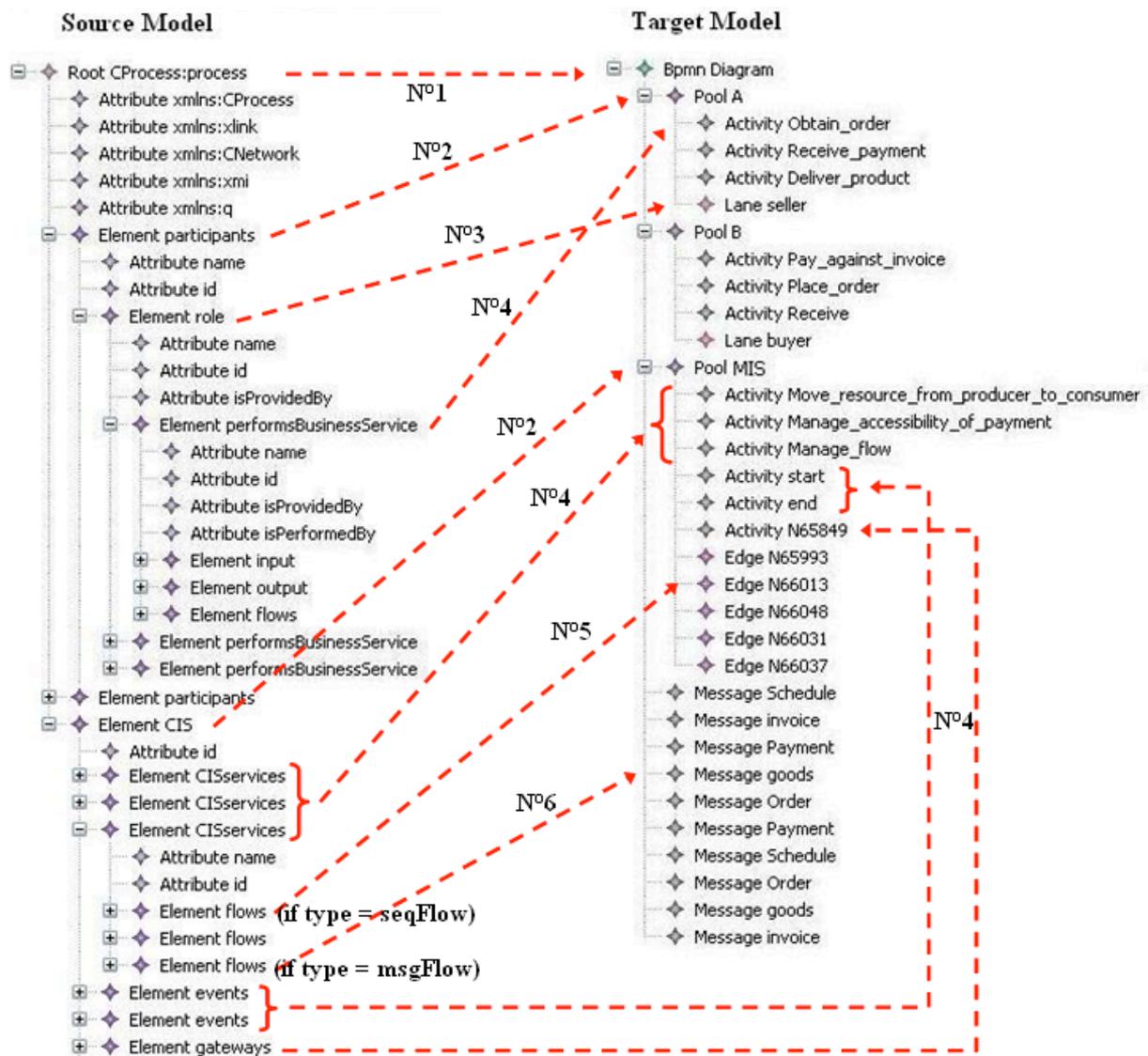

**Figure 16 Six transformations of the source model and the target model (Ecore view)**

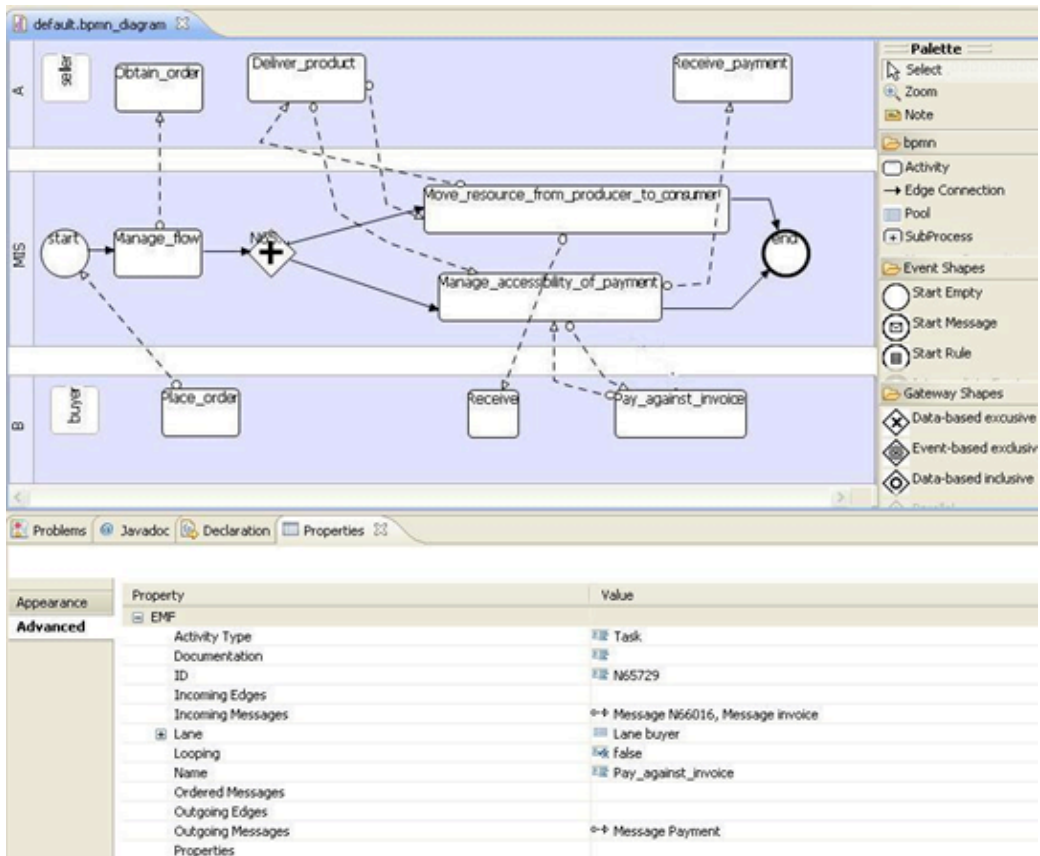

**Figure 17 Interface of the STP BPMN Modeler**

**Table**

| N° | XML source elements (NE-based) → | Result elements (OWL-based KB) |
|---|---|---|
| 1 | CNetwork:network | CNetwork |
| 2 | topology | Topology |
| 3 | commonGoals | CommonGoal |
| 4 | participants | Participant |
| 5 | role | playRole |
| 6 | relationship | Relationship |

**Table 1 XML source elements and their corresponding result elements**

| N° | XML meta-model (Source) → | BPMN meta-model (Target) |
|---|---|---|
| 1 | Root | BpmnDiagram |
| 2 | Element with the name's value *participants* or *CIS* | Pool |
| 3 | Element with the name's value *role* | Lane |
| 4 | Element with the name's value *performsBusinessService*, *CISservices*, *gateways* or *events* | Activity |
| 5 | Element with the name's value *flows* and type *seqFlow* | Edge |
| 6 | Element with the name's value *flows* and type *msgFlow* | Message |

**Table 2 Mapping between the elements of XML and BPMN meta-models**